# Network-Distributed Video Coding

Johan De Praeter, Christopher Hollmann, Rickard Sjöberg,
Glenn Van Wallendael, and Peter Lambert

***Abstract*—Nowadays, an enormous amount of videos are streamed every day to countless users, all using different devices and networks. These videos must be adapted in order to provide users with the most suitable video representation based on their device properties and current network conditions. However, the two most common techniques for video adaptation, simulcast and transcoding, represent two extremes. The former offers excellent scalability, but requires a large amount of storage, while the latter has a small storage cost, but is not scalable to many users due to the additional computing cost per requested representation. As a third, in-between approach, network-distributed video coding (NDVC) was proposed within the Moving Picture Experts Group (MPEG). The aim of NDVC is to reduce the storage cost compared to simulcast, while retaining a smaller computing cost compared to transcoding. By exploring the proposed techniques for NDVC, we show the workings of this third option for video providers to deliver their contents to their clients.**

## I. Introduction

THE way in which people consume video has evolved. In the past, video was mainly distributed to consumers through public broadcasting and physical media such as VHS tapes and DVDs. Nowadays, however, people mainly consume video through the internet by means of over-the-top (OTT) video streaming. As a result, in 2016, 73% of all consumer internet traffic consisted of video data, with forecasts indicating that this share will increase to 82% by 2021 [1].

Due to the large variety of target client devices and network conditions, real-time adaptation of the video has become a necessity. For example, a user watching a video on a smartphone with a mobile broadband connection, should receive a video with different bit rate and spatial resolution than a user streaming a video to an Ultra-High Definition television at home. As a result, different end-users will receive different versions of the same video content. These versions are referred to as representations.

J. De Praeter, G. Van Wallendael, and P. Lambert are with IDLab, imec – Ghent University, Ghent, Belgium. E-mail: (johan.depraeter@ugent.be, glenn.vanwallendael@ugent.be, peter.lambert@ugent.be).

C. Hollmann and R. Sjöberg are with Ericsson Research, Stockholm, Sweden. E-mail: (christopher.hollmann@ericsson.com, rickard.sjoberg@ericsson.com

In order to offer different video representations, providers can implement one of two main approaches. The first approach, known as simulcast, consists of encoding different independent representations of the same video and storing them on a server. When a user requests a video, the appropriate representation is selected based on device properties and current network conditions. If the network conditions change over time, the client device will switch to a more appropriate representation in real time [2]. Since each representation only has to be encoded once regardless of the number of times it is consumed, this approach scales well for systems with many users. Although this makes it a suitable approach for many video-on-demand applications, the simulcast approach has two main disadvantages. First of all, it requires a relatively large amount of storage space to store the different representations. And second, the amount of different representations is limited to a pre-determined, finite set, meaning that only a subset of the possible spatial resolutions and bit rates are covered.

If a video is not requested very often, a provider might consider it to be too costly to store many different representations. In these cases the provider can choose an alternative approach, called transcoding [3]. Contrary to simulcast, a transcoding approach requires only one high-quality representation of the video to be stored on the server. From this one representation, all other representations can be deduced. Whenever a client device requests a video, the stored representation will be adapted by the transcoder to best suit the device and bandwidth requirements of the user. However, this transcoding operation is very computationally complex. Moreover, since the transcoded representation is not stored, each new request would require a new transcoding. Since this method does not scale well, having many requests would result in high processing costs. Hence, many efforts have been made in the past to reduce the complexity of transcoding. However, the resulting complexity reductions remain relatively small.

The simulcast and transcoding approaches represent two extremes. While the former is costly in terms of storage cost, but less resource-demanding in terms of processing power, the latter is very resource-demanding in a case with many simultaneous viewers, yet much less costly in terms of storage cost. In some cases, a compromise between the two might be more desirable. For example, an early idea first proposed in 2012 suggests to store only one high-quality bitstream representation on the server, together with metadata that contains information to adapt this representation to other versions with a significantly decreased computational complexity [4]. Using such an approach, the storage cost on the server is smaller than in the case of simulcast, and the processing cost is much smaller than in the case of transcoding. Furthermore, this approach can be expanded by transmitting certain sidestreams together with the single representation to other nodes in the network, which makes it possible to generate other representations in those nodes instead of on the server. Such a method where part of the creation of other bitstream representations is offloaded to another place in the network, is called Network Distributed Video Coding (NDVC), making it a third possible approach besides simulcast and transcoding.

In July 2017, the Moving Picture Experts Group (MPEG) defined requirements [5] and released a Call for Evidence (CfE) on NDVC [6]. The goal of this call was to find middle ground between simulcast and transcoding solutions in the context of High Efficiency Video Coding (HEVC): proposals had to offer a solution that reduced the storage cost of the different representations compared to simulcast, while the computational complexity of adapting the video to the requests of clients had to be smaller than the computational complexity of transcoding.

Based on the results of the evaluation of the submitted evidence [7]–[11], MPEG concluded that the submitted technology provides benefits over simulcast and full transcoding [12]. This tutorial paper serves to make the efforts on this topic known to a broader audience.

## II. Video adaptation

In OTT video streaming, video is delivered over the public internet. Contrary to traditional broadcast television, where video is transported over dedicated networks, videos streamed over the internet will be more subject to fluctuations in data rates. These fluctuations may cause a video player to freeze if the bit rate of the video exceeds the available bandwidth of the network and no more frames are available in the buffer. In order to prevent such video freezes due to rebuffering, the video should be adapted to the available bandwidth capacity.

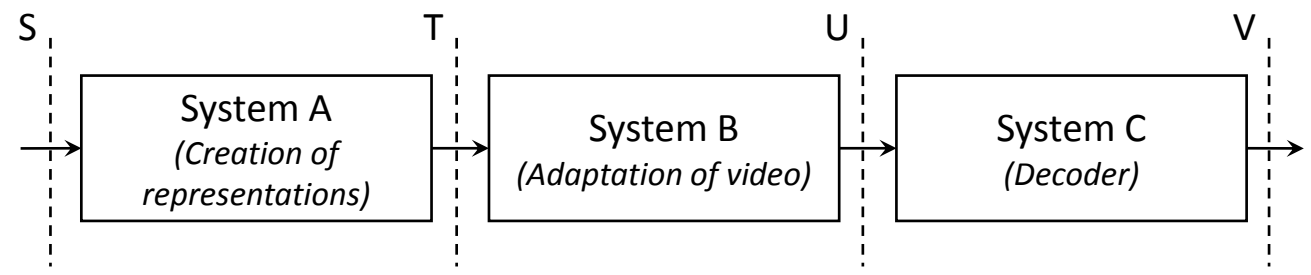

Fig. 1. Schematic overview of a general architecture of a video distribution system, with Interfaces S, T, U, and V, and Systems A, B and C.

In order to adapt a video to the available bandwidth, different representations of the video are created, with each representation having a different bit rate. Decreasing the bit rate is, for example, achieved by increasing the quantization parameter during video encoding (which removes details of the video, resulting in coding artifacts), or by downscaling the video to a smaller spatial resolution. By tuning these parameters, a series of representations ranging from high quality with high bandwidth usage to low quality with low bandwidth usage are created.

Fig. 1 presents a schematic overview onto which different techniques for video adaptation can be mapped. This architecture consists of systems that process the video and are connected to each other through generic interfaces. These interfaces are represented by the letters S, T, U, and V and can, for example, be a dedicated cable connection, public wired or mobile internet, or even an HDMI-cable going from a set-top box to a television set.

Interface S is the link over which the source video is transported to System A. In this article, this interface is assumed to always have a sufficient bandwidth capacity to transport the video. In System A, one or more representations are then created by encoding the source video. The resulting representation is then stored at Interface T. Once a user requests the representation, it is forwarded to System B, where the actual video adaptation logic is executed. The chosen representation is then transported over Interface U to System C, where it is decoded for output through Interface V. Interfaces S and V are not considered any further in this article.

In the rest of this section, three traditional approaches for video adaptation are presented by mapping them on the above architecture.

### A. Simulcast

The simulcast approach is the most widespread video adaptation approach used by content providers. This approach can be mapped on Fig. 1 as follows. System A will create $n$ different representations of the video. These representations all have different properties, such as being encoded with different bit rates, spatial resolutions, and even compression formats. When a client requests a video, the appropriate representation will be selected based on the device properties and current network conditions. In order to dynamically adapt to such changing network conditions, each representation is divided into segments of several seconds [2]. Depending on the current conditions, the selector logic in System B will choose a segment of a different representation to transmit over Interface U.

The main advantage of this method is that it scales well for systems with many users. Every representation only has to be encoded once, irrespective of the number of clients. However, this also leads to the main disadvantage of this method. For each representation, the original video has to be made available in System A and each representation has to be stored in Interface T. This can result in large bandwidth and storage cost requirements, especially if the provider wants to make many representations of many different videos available. Moreover, if a certain representation is only rarely requested by users, the storage cost may outweigh the benefits.

Another disadvantage of this method is that it is not suitable for applications that require a low latency, such as in the case of video conferencing and remote learning. Due to the use of segments, switching to a different representation is only possible at the start of a new segment. To alleviate this problem in a video-on-demand scenario, the client buffers part of the video in advance, which cannot be done in low-latency video streaming. Hence, in the latter case, the video may freeze if the bandwidth is insufficient to download the frames in the segment in time.

### *B. Just-in-time transcoding*

If storage cost is a bottleneck, a transcoding approach can be used as an alternative to simulcast. With this approach, System A creates only one high quality representation $R_0$ per video. When a different representation $R_k$ is requested by a user, System B, which contains the transcoding module, loads $R_0$ from the storage and transcodes $R_0$ to $R_k$, meaning that it is decoded, rescaled and re-encoded with different bit rate parameters. The resulting representation $R_k$ could then be temporarily cached to serve more users that request the same file.

The main advantage of this method is that only one representation is stored at Interface T, greatly reducing the storage cost compared to simulcast. Moreover, rather than simply producing a predefined representation, the transcoder can also adapt its output bit rate to match changing network conditions of the client. Since the encoder adapts dynamically, structural delays due to segment sizes as in the simulcast approach can be avoided, making this scenario ideal for low-latency use cases. However, the transcoding process is very computationally complex, and to truly provide low latency, each user would require a separate transcoder. As a result, the processing cost of this approach may become a bottleneck as more clients connect to the system.

In order to decrease the impact of the major disadvantage of transcoding, much research has focused on reducing the complexity of transcoding operations [3]. In general, such techniques extract data about coding decisions during the decoding process and use these to predict coding decisions during re-encoding. If the encoder has to evaluate less options, the computational complexity of the re-encoding step is greatly reduced. For example, for the reference encoder of the HEVC video compression standard, this complexity is often reduced by a factor of two to four. However, such faster encoding decisions come at the cost of reduced compression efficiency, meaning that such a fast transcoder may result in a bit rate overhead compared to a non-accelerated transcoder to achieve the same perceptual quality.

It should be noted that even a non-accelerated transcoder results in extra bit rate overhead compared to a simulcast scenario when generating a representation. In a simulcast scenario, a representation is directly created from the source video, whereas in transcoding, the new representation is created from a pre-existing high-quality representation. This implies that the compressed high-quality representation is decoded and re-encoded. Due to encoding always being a lossy process, this second encoding step is likely to result in a small loss of coding efficiency.

### *C. Scalable Video*

As a third possible approach besides simulcast and transcoding, video providers can also use scalable video coding to provide different representations [13]. With this technique, a single bitstream contains multiple representations. This bitstream is created in System A and consists of a low-quality base layer with multiple enhancement layers. These enhancement layers increase the spatial resolution and video quality of the base layer, with each additional layer matching the representations as in the simulcast scenario. System B will then manage which layers are dropped before the bitstream is transmitted over Interface U.

Since enhancement layers have access to the base layers, the bitstream containing all representations can be stored using less space compared to simulcast, where each representation is encoded independently. For example, storing five representations as one base layer and four enhancement layers, requires less storage than storing five simulcast representations, reducing the overall storage cost and bandwidth cost over Interface T. However, for example in the case of SHVC, the scalable version of HEVC, compression efficiency of enhancement layers compared to non-scalable HEVC decreases with up to 22% [13]. With the same example as above, it means that providing a user with a specific representation which requires both an enhancement layer and the base layer, costs 22% more bit rate on Interface U compared to providing them with the equivalent simulcast representation.

A major roadblock for the adoption of scalable video, is the fact that System C (for example the set-top box at home) must support the decoding of the enhancement layers, since only the base layer is compliant with the non-scalable version of video compression standards. As a result, this approach is rarely used by video providers, who instead opt for simulcast and/or transcoding to produce standard-compliant representations. Therefore, scalable video will not be considered further in the rest of this article.

## III. Network-Distributed Video Coding

The idea of NDVC is to have a system for video encoding and decoding where processing is distributed across three or more processing units, while the resulting bitstreams are compliant with standards supported by current hardware decoders. When mapped to the general architecture shown in Fig. 1, a simulcast approach requires a large bandwidth capacity on Interface T but little processing power in System B. Transcoding, on the other hand, results in a smaller bandwidth usage on Interface T at the cost of increased processing requirements in System B. The main goal of NDVC is to find a balance between these two extremes, thus reducing the bandwidth cost on Interface T compared to simulcast, while having a lower computational complexity in System B compared to transcoding [6].

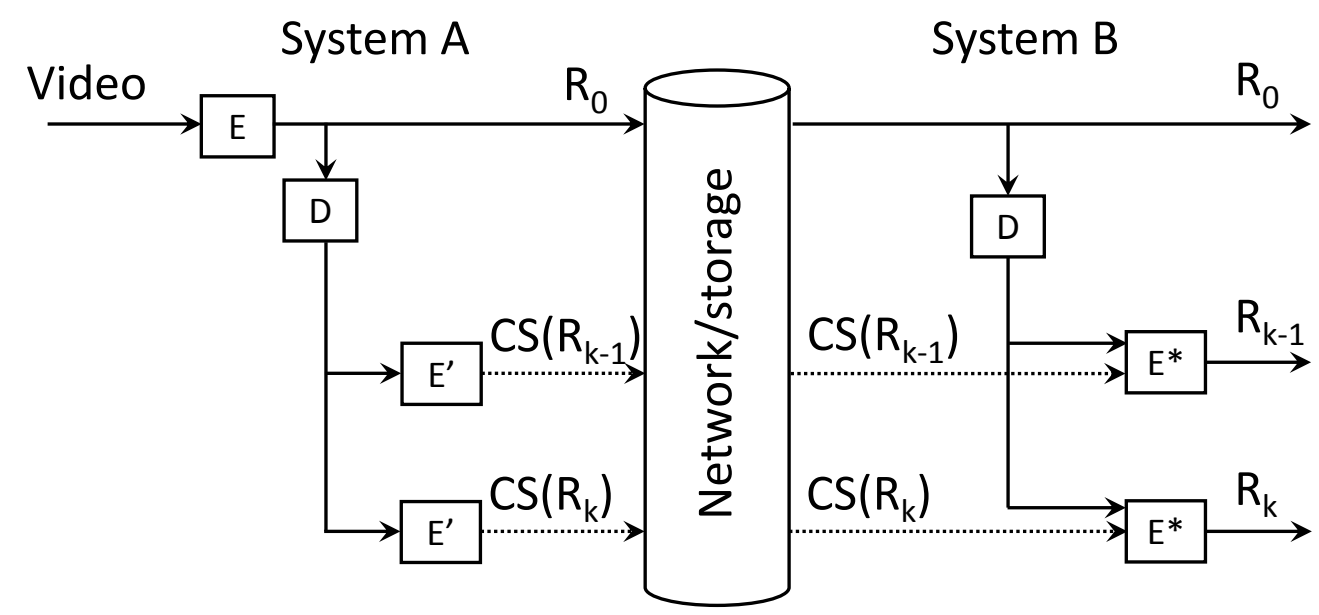

Fig. 2. Basic control stream architecture with a video encoder (E), video decoders (D), and video encoders (E') that calculate coding decisions and package these in a control stream (CS) for a certain representation $R_k$. These control streams are used as an input by fast encoders (E*) to create different representations.

### A. Basic control streams

All proposals submitted to MPEG used an approach derived from guided transcoding using control streams [4]. The basic idea of control streams is that video bitstreams consist of two major components: coding decisions and residual information. The coding decisions consist, for example, of the block structure into which a video frame is divided and the motion vectors used for inter-prediction. The goal of these coding decisions is to exploit temporal and spatial redundancy in the video by predicting the values of pixels from spatially and temporally co-located pixels. However, since this prediction is not perfect, a compensation signal known as the residual information is used to correct the prediction errors. To achieve stronger compression, this residual can be quantized with a higher value, which results in discarding more information, meaning that the bit rate of the video will be lower, but the quality will decrease as well.

When compressing a video, the encoder spends the majority of computing time to determine the coding decisions. However, these decisions contribute to less than half to the total amount of data stored in the bitstream. Instead, it is the residual information that contributes the most to the storage cost. Therefore, the basic idea of control streams is to strip a bitstream of its residual data, leaving only the coding decisions in the control stream. As a result, the storage cost is reduced by at least 50%. When requested, the bitstream can then be reconstructed by an encoder that uses coding decisions stored in the control stream to bypass all computations related to making these decisions. Consequently, the encoder will only have to recalculate the residual information, which is much less computationally complex.

This basic idea of control streams was adapted to the NDVC architecture as follows [9]. First, as seen in Fig. 2, a high-quality representation ($R_0$) of the video is created in System A. This representation is also immediately decoded at System A and re-encoded as different representations, effectively mimicking the transcoding operation that will happen later at System B. For these new representations, only the control stream is retained, whereas the residual information is discarded. The high-quality representation and the control streams are then stored. After a user requests a representation $R_k$, System B loads $R_0$ and the respective control stream $CS(R_k)$ and creates $R_k$ by transcoding $R_0$. However, contrary to the transcoding approach, the control streams will allow the transcoder to skip the coding decision calculation, resulting in a much smaller complexity compared to the full transcoding approach. Conceptually, this results in a computational cost similar to a decoder of the same video standard. Furthermore, if the same method is used to generate the residual in both System A and B, the representations on Interface U will be identical to that of a traditional transcoder.

### B. Guided transcoding for multiple representations

Although the use of control streams can greatly reduce the bandwidth cost by omitting the residual information, it still requires the creation of one control stream per representation. However, research has shown that the coding decisions of representations with the same spatial resolution resemble the decisions of representations with similar bit rates [7]. Hence, instead of using one control stream per representation, one control stream can be used for multiple representations. The inefficiencies that are introduced by using non-optimal coding decisions are then compensated by an increased amount of residual information, which means that the bit rate needed on Interface U for a certain quality may be higher than for traditional transcoding.

This idea can be applied to NDVC as follows [10]. E.g. if one spatial resolution has four different bit rate representations such as 4 Mbps, 3.1 Mbps, 2.6 Mbps, and 2 Mbps, it is only necessary to create a control stream for the representation of 2.6 Mbps. This control stream can then be used in System B to generate any of the four representations, and can even generate any rates in between using the same control stream. This is further illustrated in Fig. 3 where one control stream is generated in System A and used as an input for four encoders in System B. By reducing the amount of control streams, the necessary bandwidth on Interface T is reduced as well. However, using non-optimal coding decisions will result in extra bit rate overhead for the same quality on Interface U. This is further discussed in the comparison with simulcast and transcoding.

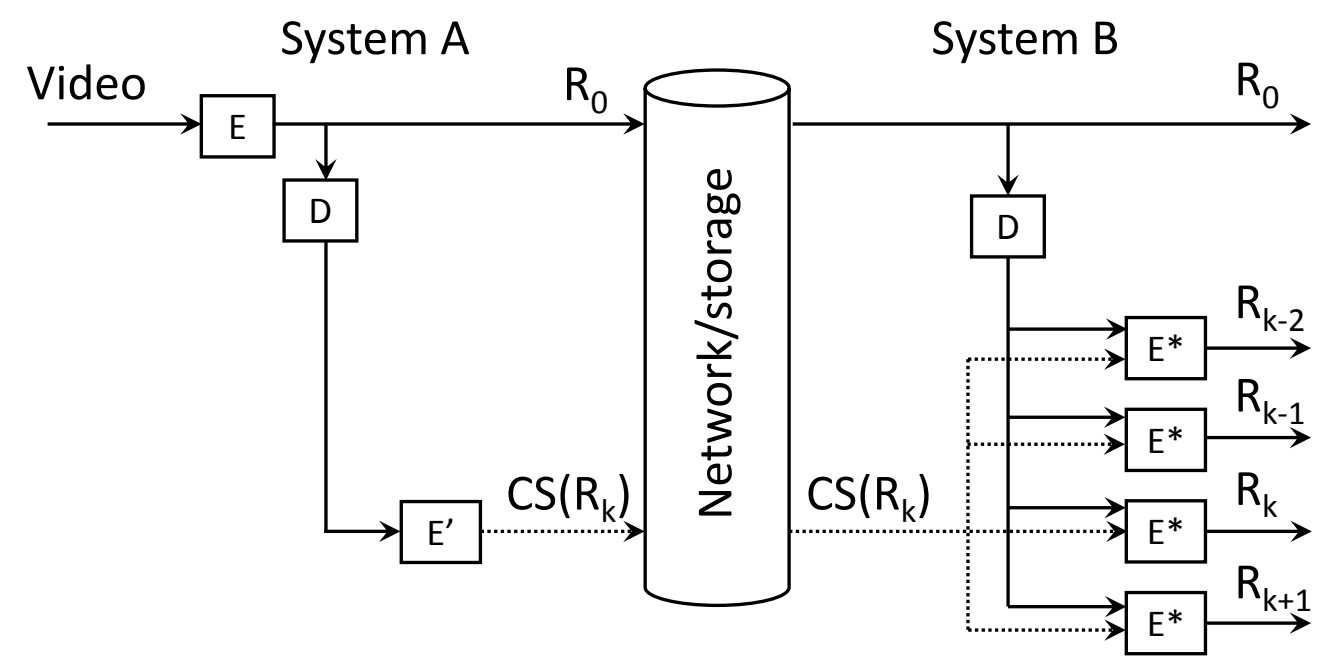

Fig. 3. Architecture with a video encoder (E), video decoders (D), and a video encoder (E') that calculates coding decisions and packages these in a control stream (CS) for a certain representation $R_k$. A single control stream is now used by a fast encoder (E*) to create multiple representations.

### C. Deflation and inflation

One way to eliminate the loss in compression efficiency stemming from transcoding is to apply a technology called deflation and inflation [8], [11]. Here, as shown in Fig. 4, the source video is used as input for the control stream generation. This is different from the basic generation of control streams shown in Fig. 3 where the decoded $R_0$ stream and not the source is used for control stream generation. The encoder E” in Fig. 4 therefore acts as a simulcast encoder. The coding decisions made by encoder E” are then applied to the decoded video of $R_0$. This generates residual coefficients which are then subtracted from the transform coefficients calculated from the source video. This subtraction results in a new set of coefficients, the so-called delta coefficients (DTC) that are included in the control stream. Once a specific representation is requested by a user, the encoder E** in System B uses the DTC to exactly reconstruct the original transform coefficients that were generated by encoder E”. The output bitstream $R_k$ after transcoding is therefore identical to the representation generated using the simulcast approach.

## IV. Comparison with simulcast and transcoding

In order to compare the proposals for NDVC to the simulcast and transcoding approaches, the test conditions defined in the call for evidence were used [6]. These conditions define three sequences of UHD resolution for which twelve representations must be created, and three sequences of HD resolution for which eight representations must be created. The test conditions also define target bit rates for the representation depending on the used sequence, with actual bit rates ranging from 10,500 kbps down to 420 kbps. All approaches were implemented based on the HEVC test model (HM) version 16.15.

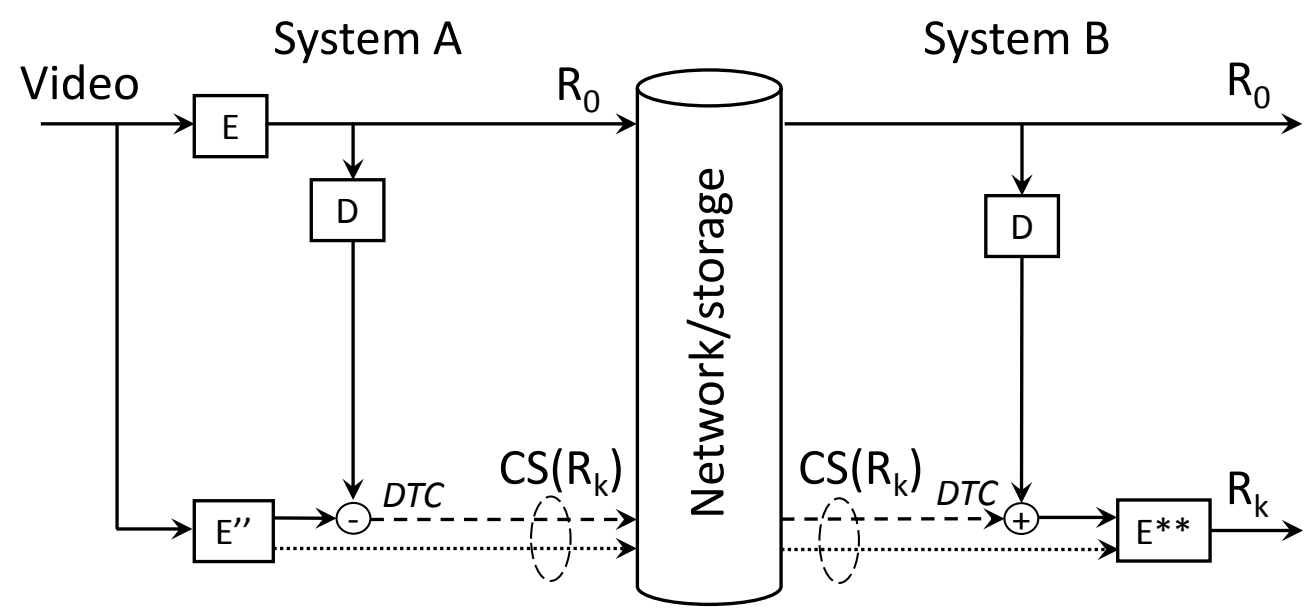

Fig. 4. Architecture for the inflation/deflation approach with a video encoder (E), video decoders (D), and video encoders that calculate coding decisions (E”) and the difference between the residual and the residual determined by applying coding decisions on the decoded high-quality representation $R_0$, the so-called delta coefficients (DTC). These are packaged together with coding decisions into a control stream (CS) for a certain representation $R_k$. These control streams are used as an input by fast encoders (E**) to perfectly reconstruct a representation with the same quality as the simulcast approach.

All variants of NDVC are compared to simulcast and transcoding in terms of cost expressed in bit rate overhead for the same quality on interfaces T and U, as well as in terms of processing time compared to full transcoding. The evaluated configurations are shown in Table I. NDVC was evaluated for a configuration using deflation and inflation with one control stream per representation, as well as for configurations in which respectively one, two, and four representations are created using the same control stream.

Deflation/inflation manages to reduce the cost on Interface T compared to simulcast with 27.1%, with a complexity reduction of 99.7% compared to full transcoding using the reference encoder. Moreover, the cost on Interface U remains identical to simulcast.

Without using deflation/inflation, the cost on Interface T is reduced by 50.5% up to 68% depending on the amount of representations per control stream. This cost reduction comes much closer to the one of full transcoding, while having a much smaller computational complexity. Although using one representation per control stream results in the same cost on Interface U as transcoding, using a smaller amount of control stream for the same amount of representations results in an increasing cost on Interface U.

When comparing the different variants of NDVC, it becomes clear that whereas all variants greatly reduce the computational complexity compared to full transcoding, a smaller cost on Interface T results in a larger cost on Interface U. These results show that NDVC enables a middle way between simulcast and transcoding, by offering content providers a way to make a trade-off based on the available bandwidth on Interfaces T and U.

## V. Conclusion

This article presented the concept of NDVC as an alternative to the traditional approaches of simulcast and transcoding for video adaptation. This approach focuses on distributing the processing of video representations across three or more different processing units in the network, while allowing a trade-off in bandwidth usage on the interfaces between the units. In particular, we have focused on the techniques discussed at MPEG, which either offer the same quality as the simulcast solution, while having a lower storage cost, or offer a similar quality as transcoding, at a much lower processing cost.

For video providers, the techniques of NDVC offer a solution to reduce the storage costs of video representations that are requested less frequently, while reducing the computational complexity by more than 98% compared to full transcoding using the reference encoder. From a user perspective, if they wish to view different representations of a video (e.g. on different devices) and the highest quality representation was already downloaded on their set-top box, they only need to download the necessary metadata instead of downloading the new representation in its entirety. Furthermore, by having access to low-complexity transcoding close to the user, in real-time video streaming scenarios, the bit rate can be adapted quickly. Finally, contrary to scalable video compression standards such as SHVC, NDVC does not require additional hardware for end-user devices, since the resulting streams are standard-compliant with common hardware decoders for current video compression standards such as H.264/AVC and HEVC.

## References

[1] Cisco, “Cisco Visual Networking Index: Forecast and Methodology, 2016-2021,” June 2017, [Accessed Nov. 23, 2018]. [Online]. Available: https://www.cisco.com/c/en/us/solutions/collateral/service-provider/visual-networking-index-vni/complete-white-paper-c11-481360.pdf

TABLE I
COMPARISON BETWEEN SIMULCAST, NDVC WITH DEFLATION/INFLATION, NDVC WITH RESPECTIVELY ONE, TWO, AND FOUR REPRESENTATIONS PER CONTROL STREAM (R/CS), AND TRANSCODING.

| | | NDVC | | | | |
|---|---|---|---|---|---|---|
| | Simulcast | Deflation/inflation | 1 R/CS | 2 R/CS | 4 R/CS | Full transcoding |
| Cost (%) vs simulcast (T) | 0.0 | -27.1 | -50.5 | -64.8 | -68.0 | -74.6 |
| Cost (%) vs simulcast (U) | 0.0 | 0.0 | 8.5 | 13.9 | 18.5 | 8.5 |
| Cost (%) vs full transcoding (T) | 297.1 | 189.5 | 96.7 | 39.6 | 26.4 | 0.0 |
| Cost (%) vs full transcoding (U) | -6.4 | -6.4 | 0.0 | 6.2 | 10.8 | 0.0 |
| Time (%) vs full transcoding (B) | -100.0 | -99.7 | -98.6 | -98.3 | -98.1 | 0.0 |

[2] I. Sodagar, “The MPEG-DASH Standard for Multimedia Streaming Over the Internet,” *IEEE Multimedia*, vol. 18, no. 4, pp. 62–67, Apr. 2011.

[3] A. Vetro, C. Christopoulos, and H. Sun, “Video transcoding architectures and techniques: an overview,” *IEEE Signal Process. Mag.*, vol. 20, no. 2, pp. 18–29, Mar. 2003.

[4] G. Van Wallendael, J. De Cock, and R. Van de Walle, “Fast transcoding for video delivery by means of a control stream,” in *Proc. IEEE Int. Conf. Image Process. (ICIP)*, Sept. 2012, pp. 733–736.

[5] R. Sjöberg, X. Ducloux, K. Park, and R. Mekuria, “Requirements for network distributed video coding (version 5),” ISO/IEC JTC1/SC29/WG11, Tech. Rep. MPEG2017/N17063, July 2017.

[6] “Call for Evidence on Transcoding for Network Distributed Video Coding,” ISO/IEC JTC1/SC29/WG11, Tech. Rep. MPEG2017/N17058, July 2017.

[7] J. De Praeter, G. Van Wallendael, J. Slowack, and P. Lambert, “Video encoder architecture for low-delay live-streaming events,” *IEEE Trans. Multimedia*, vol. 19, no. 10, pp. 2252–2266, Oct. 2017.

[8] C. Hollmann and R. Sjöberg, “Guided transcoding using deflation and inflation,” in *Proceedings of the 23rd Packet Video Workshop*. ACM, June 2018, pp. 19–24.

[9] J. De Praeter, G. Van Wallendael, and P. Lambert, “CfE NDVC: Guided transcoding using CIC and RE modules,” ISO/IEC JTC1/SC29/WG11, Tech. Rep. MPEG2017/m41825, Oct. 2017.

[10] ——, “Extension on guided transcoding using CIC and RE modules in the context of CfE NDVC,” ISO/IEC JTC1/SC29/WG11, Tech. Rep. MPEG2017/m41826, Oct. 2017.

[11] C. Hollmann, R. Sjöberg, and K. Andersson, “NDVC: CfE response from Ericsson,” ISO/IEC JTC1/SC29/WG11, Tech. Rep. MPEG2017/m41829, Oct. 2017.

[12] R. Sjöberg, G. Van Wallendael, and X. Ducloux, “Report on the Call for Evidence on Transcoding for Network Distributed Video Coding,” ISO/IEC JTC1/SC29/WG11, Tech. Rep. N17261, Oct. 2017.

[13] J. Nightingale, Q. Wang, C. Grecos, and S. Goma, “Video adaptation for consumer devices: opportunities and challenges offered by new standards,” *IEEE Commun. Mag.*, vol. 52, no. 12, pp. 157–163, Dec. 2014.

**Johan De Praeter** received his M.Sc. and Ph.D. degree in Computer Science Engineering from Ghent University, Ghent, Belgium, in 2013 and 2017, respectively. Since 2013, he has been with IDLab, Ghent University – imec, where he is currently working as a post-doctoral researcher on the topic of multimedia processing. His main research interests include computational complexity reduction of video compression, video transcoding, and network distributed video coding.

**Christopher Hollmann** obtained his M.Sc. degree from Uppsala University, Sweden, in 2017. In 2016 he joined Ericsson Research in Stockholm, Sweden, as a Researcher in the Visual Technologies group. His research interests include network distributed video coding and video compression technologies.

**Rickard Sjöberg** received the M.Sc. degree in computer science in 1997 from the Royal Institute of Technology, Stockholm, Sweden. He has been with Ericsson since 1996 and has worked in various areas related to video coding, in both research and product development. In parallel, he has been an active contributor in the video coding standardization community, with hundreds of contributions to the video coding standards of ITU-T, MPEG, JVT, JCT-VC and JVET. He is currently working as a technical lead of video coding research at Ericsson Research in Stockholm, Sweden.

**Glenn Van Wallendael** obtained the M.Sc. degree in Applied Engineering from the University College of Antwerp, Belgium, in 2006 and the M.Sc. degree in Engineering from Ghent University, Belgium in 2008. Afterwards, he worked towards a Ph.D. at IDLab, Ghent University - imec, with the financial support of the Research Foundation – Flanders (FWO), obtaining it in 2013. Currently, he continues working in the same group as a post-doctoral researcher. His main topics of interest are video compression, including scalable video compression, and transcoding.

**Peter Lambert** is a full-time Associate Professor at IDLab, Ghent University – imec (Belgium). He received his Master’s degree in Science (Mathematics) and in Applied Informatics from Ghent University in 2001 and 2002, respectively, and he obtained the Ph.D. degree in Computer Science in 2007 at the same university. In 2009, he became a Technology Developer at Ghent University, which he combined with a part-time Assistant Professorship at IDLab since 2010 before becoming full-time Associate Professor in 2013. His research interests include (mobile) multimedia applications, multimedia coding and adaptation technologies, and 3D graphics.